\documentclass[11pt,a4paper]{article}
\usepackage[T1]{fontenc}
\usepackage[utf8]{inputenc}
\usepackage{authblk}
\usepackage{url}
\usepackage{graphicx}
\usepackage[margin=1.2in]{geometry}
\usepackage{color}
\usepackage{amsmath}
\usepackage{pdflscape}
\usepackage{float}
\usepackage{subcaption}
\newcommand{\s}[1]{{{#1}}}

\title{Data Mining of Online Genealogy Datasets for Revealing Lifespan Patterns in Human Population}
\author{Michael Fire and Yuval Elovici\thanks{Email:\{mickyfi,elovici\}@bgu.ac.il}}

\affil{Telekom Innovation Laboratories  at Ben-Gurion University of the Negev \\
Department of Information Systems Engineering, Ben-Gurion University}

\date{January 5, 2014}

\begin{document}
\maketitle

\begin{abstract}
Online genealogy datasets contain extensive information about millions of people and their past and present family connections. This vast amount of data can assist in identifying various patterns in human population.
In this study, we present methods and algorithms which can assist in identifying variations in lifespan distributions of human population in the past centuries, in detecting social and genetic features which correlate with human lifespan, and in constructing predictive models of human lifespan based on various features which can easily be extracted from genealogy datasets.

We  have evaluated the presented methods and algorithms on a large online genealogy dataset with over a million profiles and over \s{9} million connections, all of which were collected from the WikiTree website.
Our findings indicate that significant but small positive correlations exist between the parents' lifespan and their children's lifespan.  Additionally, we found slightly higher  and significant correlations between the lifespans of spouses. We also discovered a very small positive and significant correlation between longevity and reproductive success in males, and a small and significant negative correlation between longevity and reproductive success in females. Moreover, our machine learning algorithms presented better than random classification results in predicting which people who outlive the age of 50 will also outlive the age of 80.

We believe that this study will be the first of many studies which utilize the wealth of data on human populations, existing in online genealogy datasets, to better understand factors which influence human lifespan. Understanding these factors can assist scientists in providing solutions for successful aging.
\\\\
\noindent \textbf{Keywords. Genealogy Data Mining, Aging, Gerontology, Human Population Lifespan, Lifespan Prediction, Date Mining, Machine Learning, WikiTree} 

\end{abstract}

\section{Introduction}
In the last decade, Web 2.0 websites, such as Wikipedia\footnote{\url{ http://en.wikipedia.org}} and Reddit,\footnote{\url{ http://www.reddit.com}} have become extremely popular and widespread. Web 2.0 websites offer Internet users opportunities to connect, collaborate, and share information with each other, creating massive datasets with millions of content items. For example, the free encyclopedia Wikipedia has more than 4.3 million articles and more  than 127,000 active users who contribute new content to the site on a regular basis~\cite{wikipedia_stat}. One type of Web 2.0 site which recently became popular is genealogy websites.  Genealogy websites, such as MyHeritage,\footnote{\url{http://en.wikipedia.org}} Ancestry,\footnote{\url{ http://www.ancestry.com}} WikiTree,\footnote{\url{http://www.wikitree.com}} and Familypedia,\footnote{\url{ http://familypedia.wikia.com}} have millions of users~\cite{myheritage_map,ancestry_report} that use these websites to create, discover, and share their family history by generating and updating online family trees. These online family trees consist of personal data on family members from the last several centuries, and they give many personal details for each family member, such as the member's date of birth, date of death, ancestors' details, and children's details, among others.

The family tree structure and the family members' personal details that are stored in these genealogy websites create large-scale datasets, which contain billions of entries~\cite{myheritage_billion,ancestry_report} on human life and death properties. These datasets can be utilized to reveal interesting patterns regarding lifespan changes over the centuries. Additionally, these datasets can also assist in better understanding and identifying characteristics which are correlated with human lifespan changes. For example, these datasets can be explored and utilized to answer the following questions: \textit{Does having more children extend one's lifespan?} \textit{Does having long-lived ancestors prolong life?}  \textit{Does getting married lengthen one's lifespan?}
Answering these types of questions can assist scientists in providing insights and solutions for successful aging.

In this study, we present data mining algorithms for analyzing large genealogy datasets in order to examine human population lifespan variations over a substantial length of time (see Section~\ref{sec:variations}). Moreover, we introduce methods to utilize these types of datasets to identify features which correlate with human lifespan (see Section~\ref{sec:linear}).  Additionally, we also present Machine Learning (ML) algorithms based on features extracted from genealogy datasets, which can assist in predicting if a particular 50-year-old individual will reach the age of 80  (see Section~\ref{sec:ml}).

To test and evaluate our algorithms, we developed a web crawler which crawled and parsed public profiles from the WikiTree website. WikiTree is a free, collaborative family-history website, which contains more than 5 million user-contributed profiles~\cite{wikitree5m} of individuals who have lived in the past centuries, and many of the profiles contain personal details about each individual. Using the collected data from WikiTree, we were able to construct a dataset (referred to as the WikiTree dataset) of over a million \s{public} profiles, out of which at least \s{416,030} profiles were of individuals who were born in the United States (see Section~\ref{sec:wiki}). 

By analyzing the WikiTree dataset, we calculated various statistics on variations of population lifespan over the last centuries, including specific statistics on the lifespan variations of the United States population (see Section~\ref{sec:stat} and Figures~\ref{fig:lifespan_avg} and~\ref{fig:lifespan_med}). As a result of this analysis, we discovered several interesting historical lifespan change patterns (see Section~\ref{sec:discussion}); for example, we discovered that the average lifespan of females who were born in the United States and lived beyond the age of ten increased sharply in just a half-century: from \s{62.66} in  1850  to \s{72.5} in 1900  (see Figure~\ref{fig:lifespan_avg}).

Using the WikiTree dataset, we constructed a social network directed multigraph which contains over \s{1.38} million vertices and over \s{9.19} million links  (see Section~\ref{sec:construction} and Table~\ref{tab:wikitree_stat}). We then analyzed the social network graph and extracted 21 features, such as parents' and grandparents' ages of death, for each vertex in the graph (see Section~\ref{sec:features}). By using the extracted features and simple linear regression models, we discovered significant correlations with low coefficients of determination between the individuals' ages of death and the ages of death of their siblings, parents, spouses, and grandparents (see Table~\ref{tab:linear_results}). 
We also discovered a slighter higher significant correlation between the individual's age of death and the age of death of his or her spouse (see Table~\ref{tab:linear_results}). Additionally, we constructed multiple linear regression models for predicting an individual's age of death based on various features which were extracted from the individual's personal details. Our multiple linear models were with high significance and Multiple Adjusted R-squared values up to 0.085 (see Table~\ref{tab:multi_linear_results}).

Our ML classifiers have presented better than random results in predicting which individuals who outlived the age of fifty and passed the age of menopause will also outlive the age of 80 (see Section~\ref{sec:ml_results}).

The remainder of the paper is structured as follows: In Section~\ref{sec:related_work} we give a brief overview of previous relevant studies on characteristics which were found to be correlated with human lifespan. In this section, we also introduce several studies which used similar data mining  algorithms as this study. Next, in Section~\ref{sec:methods} we present the methods and algorithms we developed for studying genealogy datasets. In this section, we also describe our constructed WikiTree dataset. Then, in Section~\ref{sec:results} we present our algorithm evaluations results on the WikiTree dataset.  Lastly, in Section~\ref{sec:discussion} we discuss our results, and we also offer future research directions.

\section{Related Work}
\label{sec:related_work}
The factors that influence human lifespan have been thoroughly studied over the past decades~\cite{thomas2000human,mitchell2001heritability,mcardle2006does,le2007does,gogele2011heritability}. In this section we give a brief overview of recent genealogical studies that are most relevant to this study, pinpointing similar factors. Additionally, we also give a short overview of recent studies in the field of social network analysis and data mining, which used a similar methodology to the one used throughout this study. 

In recent years, many studies have tried to find correlations between parents' and childrens' lifespans, as well as correlations between lifespans of parents and their number of children:  
In 1998, Westendorp and Kirkwood used a historical dataset, from the British aristocracy, to study the connection between longevity and reproductive success. They discovered that longevity was positively correlated with age at first childbirth, and negatively correlated with number of children.  In 2000, Thomas et al.~\cite{thomas2000human} studied the connection between longevity and fertility using a statistical dataset of 153 countries.  They concluded that ``humans who invest heavily in reproduction while young will, on average, pay for this reproductive success with a shortened lifespan.'' In 2001, Mitchell et al.~\cite{mitchell2001heritability} used  genealogical data  of Old Order Amish members to estimate the parent-child correlations in lifespan. They also  estimated the child age of  death as a function of parent age at death. They discovered significant but small correlations between parental and child ages at death.

In 2006, McArdle et al.~\cite{mcardle2006does} studied the correlation between the number of children and lifespan using genealogical data of 2,015 individuals who were members of an Old Order Amish community. In their study they discovered lifespans of fathers increased linearly with increasing number of children, while lifespans of mothers increased linearly up to 14 children but decreased with each additional child beyond 14. 
In 2007, Le Bourg~\cite{le2007does} presented a thorough review of studies which researched the  relationship between fertility and longevity under various conditions. According to Le Bourg, the review results indicated that ``in natural fertility conditions longevity does not decrease when the number of children increases but, in modern populations, mortality could slightly increase when women have more than ca 5 children.'' 
In 2011, G\"{o}gele et al.~\cite{gogele2011heritability} conducted a comprehensive genealogical study with a thorough assessment of the heritability of lifespan and longevity in three villages in Italy. Their research, which included studying more than 50,000 individuals across four centuries, discovered ``a general low inheritance of human lifespan, but which increases substantially when considering long-living individuals, and a common genetic background of lifespan and reproduction.'' 

Many studies found connections between an excess in mortality and bereavement, also known as  the ``widow effect.'' In 1969, Parkes et al.~\cite{parkes1969broken} followed 4,486 widowers at the age of 55 for nine years. Out of these widowers, 213 died during
their first six months of bereavement, 40\% above the expected rate for married men of the same age.
In 1996, Martikainen et al.~\cite{martikainen1996mortality} conducted a large scale study of 1,580,000 married Finnish individuals and also discovered excess mortality among the bereaved.
In 2008, Elwert and Christakis~\cite{elwert2008effect} studied 373,189 elderly married couples in the United States. They discovered that the death of a spouse from almost all causes increased the mortality of the bereaved partner to varying degrees.

In our research we used several regression and ML techniques for lifespan prediction. 
In order to carry out our work, we mainly used attributes which could be extracted from genealogy datasets in order to construct the genealogy social network and extract features from the network (see Section~\ref{sec:methods}). 
Similar techniques that involve social network analysis and regression were used by Christakis~\cite{christakis2007spread} in researching the spread of obesity, by Altshuler et al.~\cite{altshuler2012incremental} in predicting the individual parameters and social links of smart-phone users, and by Fire et al.~\cite{fire2012predicting} in predicting students' final exam scores. 

\section{Methods and Experiments}
\label{sec:methods}

To cope with the challenge of analyzing a huge online genealogy dataset with ten of millions of records on individuals' personal data and their connections, we first chose to convert the dataset into a social network represented by a directed multigraph where vertices represent people and links represent connections among family members (see Section~\ref{sec:construction}). Next, we used the constructed social network graph and extracted various features from each vertex, such as the vertex's number of children, year of birth, and gender (see Section~\ref{sec:features}). We then used the extracted features to determine various statistics on the population lifespan variations over time (see Section~\ref{sec:variations}). After that, we used linear regression to find the features that significantly influence human lifespan. We also constructed multi-linear regression models for lifespan prediction (see Section~\ref{sec:linear}). Lastly, we used ML algorithms to construct classifiers which can predict if a person from the United States who outlives the age of fifty will also reach the age of eighty  (see Section~\ref{sec:ml}).

To perform our statistical calculations and to construct our predictive models, we used various datasets that were  extracted from a large genealogy dataset. These datasets are defined in Table~\ref{tab:dataset_table}. Additionally,  the methods and algorithms we have used throughout this study are summarized in Table~\ref{tab:methods_table}. 

\begin{landscape}
\begin{table}[htb]
\centering
\caption{Dataset Definitions}{ 
\begin{center}
\includegraphics[width=1.7\textwidth]{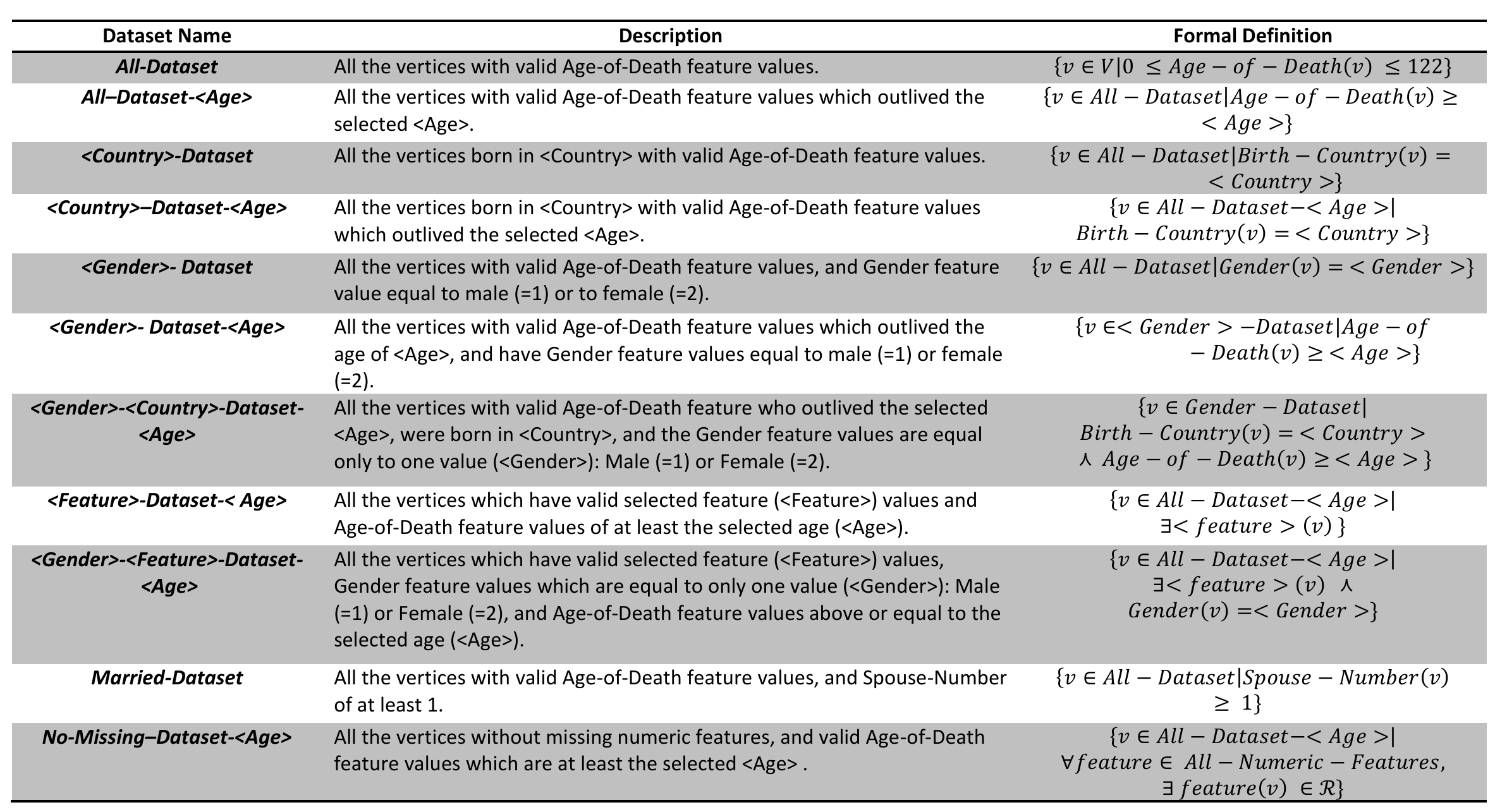}
\end{center}
}
\label{tab:dataset_table}
\end{table}

\begin{table}[htb]
\centering
\caption{Method and Algorithm Overview}{ 
\begin{center}
\includegraphics[width=1.7\textwidth]{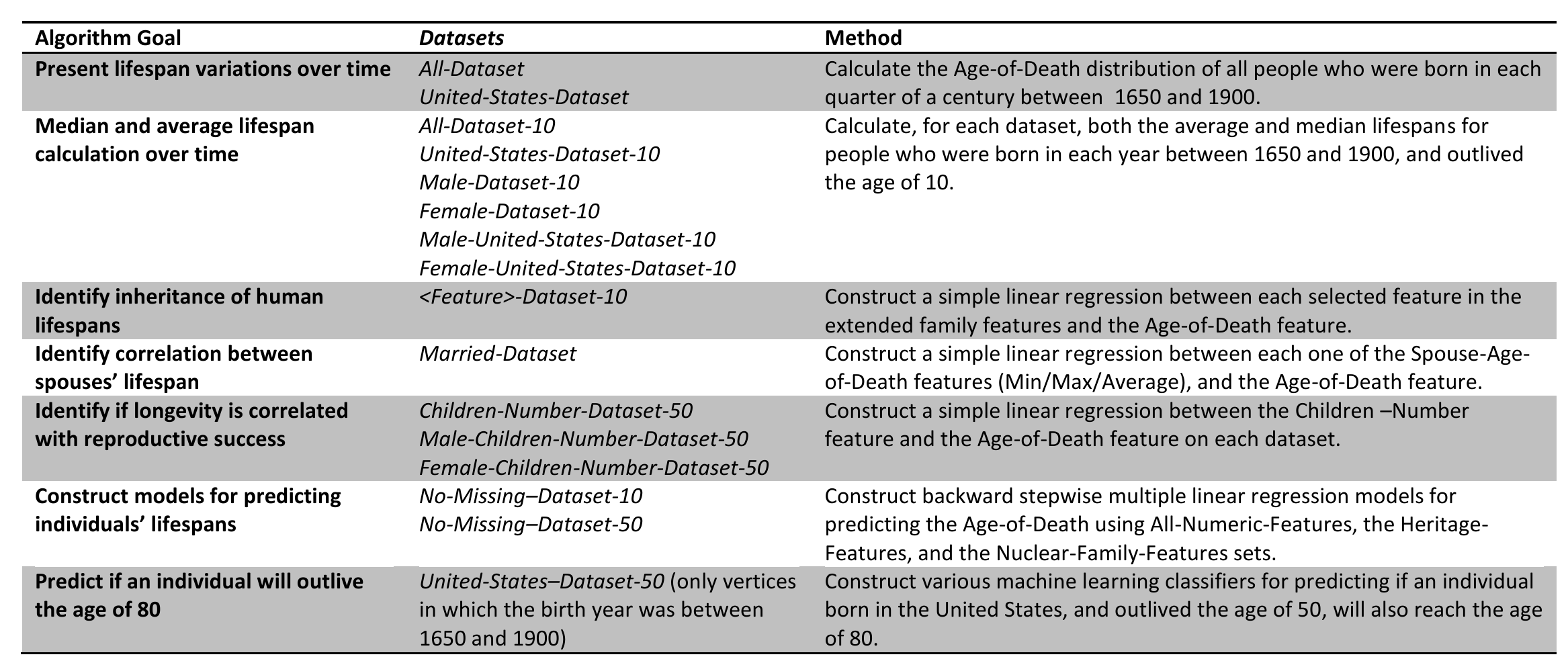}
\end{center}
}
\label{tab:methods_table}
\end{table}
\end{landscape}

\subsection{Constructing the Genealogy Social Networks}
\label{sec:construction}
We constructed the social network directed multigraph $G:=<V,E>$ from the genealogy dataset in the following manner: First, we  assembled the graph vertices set $V$ by adding a new vertex $v \in V$ for each profile in the genealogy dataset. We then defined $E$ as the multiset of links in the graph, with each link $e \in E$ defined to be a tuple $e=(u,v,t,c)\in E$, where $u,v \in V$; $t$ is the link type, which can be one of the following values: 
$t \in\{\mbox{Spouse},\mbox{Child},\mbox{Parent}, \mbox{Sibling}\}$; and $c$ is the creation date of the link. For example, if the genealogy dataset contains the profiles of Queen Elizabeth II and Prince Charles,  then the social network graph will contain the following vertices: $\mbox{Elizabeth II}, \mbox{Prince Charles} \in V$, and  the following edges: $(\mbox{Elizabeth II}, \mbox{Prince Charles}, \mbox{Parent}, \mbox{14 November 1948})$ and  $(\mbox{Prince Charles},\mbox{Elizabeth II}, \mbox{Child}, \mbox{14 November 1948})$. \s{In case the genealogy dataset contains link of type $t$ between a public profile $u$ and a private profile, we added to the multigraph a new vertex $v_{private}$ to $V$ and added new link $e=(u,v_{private},t, \emptyset)$ to $E$.}\footnote{\s{During this study, we have utilized private profiles to calculate public profiles' features, such as Children-Number($u$) and Spouse-Number($u$) more accurately (see Section~\ref{sec:features}). In many cases, we cannot distinguish if two  or more private profiles are in fact represent the same single profile in the genealogy dataset. Nevertheless, we can estimate the number of distinct private profiles by utilizing the private profile single link of type $t$. Namely, due to the fact that most people have two parents, we can conclude that $n$ private profiles with single link of  type ``Child'' represent at least $\frac{n}{2}$ distinct profiles. }}
Next, for each \s{non-private} vertex $v \in V$ in the network, we added attributes based on the information extracted from each individual's profile page, which is represented by $v$.\footnote{In most online genealogy websites, a profile usually contains the following information about each individual: gender, birth and death dates, location of birth, location of death, parents' names, spouses' names, siblings' names, and children's names.} 
Then, for each vertex  $v$, we calculated a list of features described in the following subsection.
Lastly, we \s{removed the birthday and death date data of any vertices in the graph} with the following inconsistencies in their data: (a) profiles with a negative age of death,  which can occur from reversing the birth and death dates; (b) profiles shown to be over the age of 122, the maximum recorded age~\cite{maxlifespan}; and (c) profiles shown as having children and also an age of less than five.\footnote{ The youngest mother on record was a 5-year-old Peruvian girl~\cite{murdock1998teenage}.} 

\begin{figure}[t]
\begin{center}
\includegraphics[]{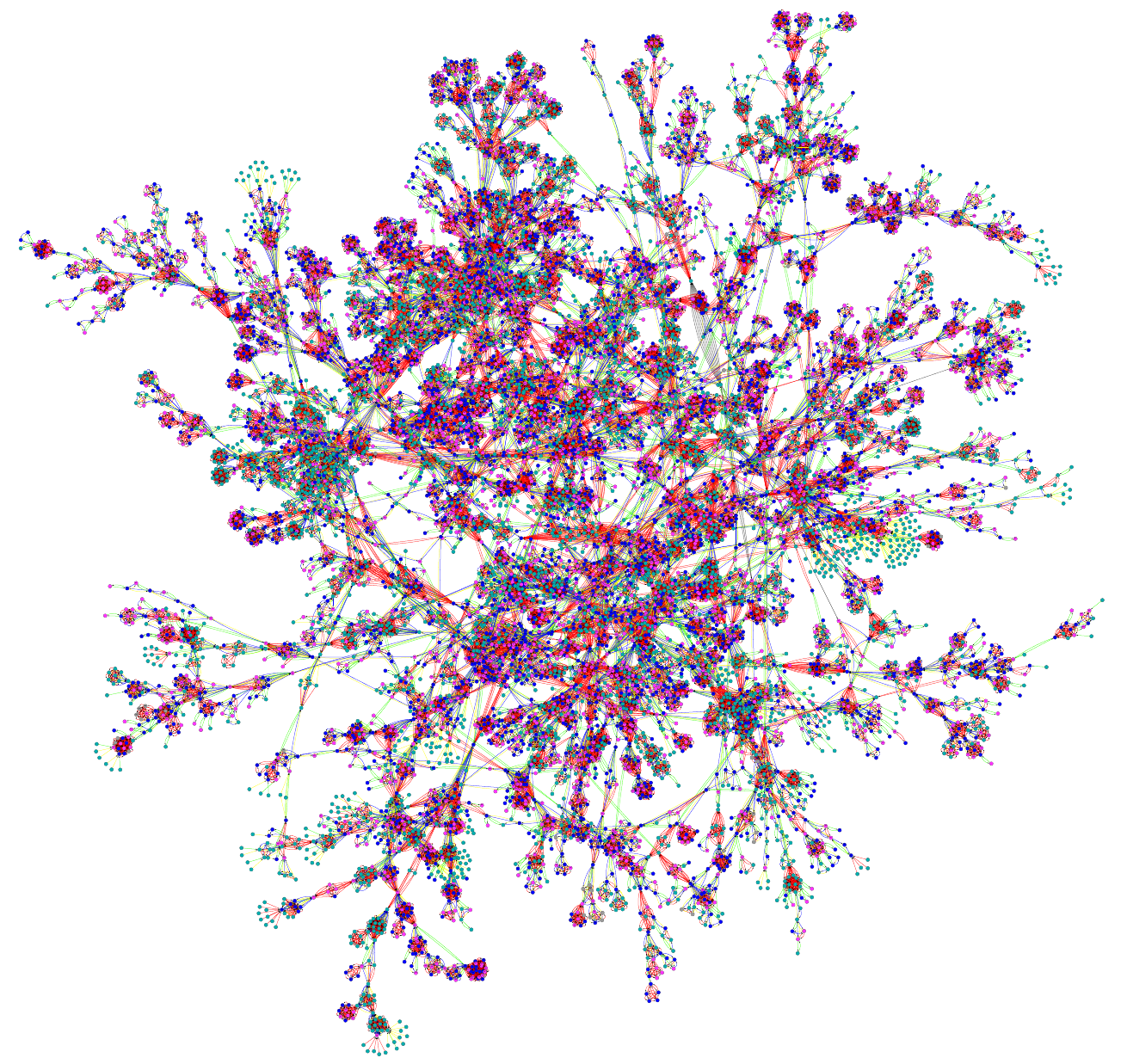}
\end{center}
\caption{
\textbf{WikiTree's directed multigraph (subgraph with 10,000 vertices).}
Each link color represents a connection of a different type: \textit{blue} is a Parent link, \textit{red} is a Sibling link, \textit{green} is a Spouse link, and \textit{yellow} is a Child link. The color of each vertex defines the vertex's gender: \textit{blue} represents male vertices, \textit{pink} represents female vertices, and \textit{gray} represents unknown gender. Each vertex label, which is visible by zooming into the graph, contains the vertex lifespan if one exists.
}
\label{fig:multi_graph}
\end{figure}

\subsection{Feature Extraction}
\label{sec:features}
After constructing the social network graph, we can extract, if possible,  three types of features for each vertex: The first type is the vertex \textit{general profile features}, which include basic information about the vertex, such as birth year, gender, and full name. The second type is the \textit{nuclear family features}, which include information about the vertex's children and spouses. The third and last type of features is the \textit{extended family features}, which include information about the vertex's parents, siblings, and grandparents. 
In this study, we extracted a total of 21 features for each vertex $v \in V$. 
In the remainder of this section, we introduce and give formal definitions for each one of these features.

\subsubsection{General Features}
\label{sec:general_features}
\begin{enumerate}
\item \textbf{Full-Name($v$)} - The full name of $v$.
\item \textbf{Birth-Year($v$)} - The birth year of $v$.
\item \textbf{Death-Year($v$)} - The death year of $v$.
\item \textbf{Gender($v$)} - The gender of $v$ converted to an integer, where male is set to 1, female is set to 2, and unknown gender is set to 0.
\item \textbf{Birth-Country($v$)} - The country in which $v$ was born.
\item \textbf{Death-Country($v$)} - The country  in which $v$ died.
\item \textbf{Age-of-Death($v$)} - The age of death of $v$ (also referred to as the lifespan of $v$) which is calculated, if accurate dates are available, by subtracting the birth date of $v$ from the death date of $v$.

\newcounter{enumTemp}
\setcounter{enumTemp}{\theenumi}
\end{enumerate}

\subsubsection{Nuclear Family Features} 
\label{sec:nuclear_features}
\begin{enumerate}
 \setcounter{enumi}{\theenumTemp}
\item \textbf{Children-Number($v$)} - the number of children which $v$ had. The formal Children-Number($v$) definition is:
\[\mbox{Children-Number}(v) := |\{u \in V|u \in V \land  \exists (v,u,Child, t) \in E\}|.\]

\item \textbf{Spouse-Number($v$)} - the number of individuals to which $v$ was married to. The formal Spouse-Number($v$) definition is:
\[\mbox{Spouse-Number}(v) := |\{u \in V|(u \in V \land  \exists (v,u,Spouse, t) \in E\}|.\]

\item \textbf{Min-Spouse-Age-of-Death($v$)} - the minimum age of death of $v$'s spouses. The formal Min-Spouse-Age-of-Death($v$) definition is:
\begin{multline*}
\mbox{Min-Spouse-Age-of-Death}(v) := \\ min(\{\mbox{Age-of-Death}(u) \neq \emptyset | u \in V \land  \exists(v,u,Spouse,t) \in E \}),
\end{multline*}
where the function \textit{min($S$)} returns the minimum value among set $S$ members, or 0 if $S$ is empty.

\item \textbf{Max-Spouse-Age-of-Death($v$)} - the maximum age of death of $v$'s spouses. The formal Max-Spouse-Age-of-Death($v$) definition is:
\begin{multline*}
\mbox{Max-Spouse-Age-of-Death}(v) := \\ max(\{\mbox{Age-of-Death}(u) \neq \emptyset | u \in V \land  \exists(v,u,Spouse,t) \in E \}),
\end{multline*}
where the function \textit{max($S$)} returns the maximum value among set $S$ members, or 0 if $S$ is empty.

\item \textbf{Avg-Spouse-Age-of-Death($v$)} - the average age of death of $v$'s spouses. The formal Avg-Spouse-Age-of-Death($v$) definition is:
\begin{multline*}
\mbox{Avg-Spouse-Age-of-Death}(v) := \\ avg(\{\mbox{Age-of-Death}(u) \neq \emptyset | u \in V \land  \exists(v,u,Spouse,t) \in E \}),
\end{multline*}
where the function \textit{avg($S$)} returns the average value of set $S$ members, or 0 if $S$ is empty.

\setcounter{enumTemp}{\theenumi}
\end{enumerate}

\subsubsection{Extended Family Features}
\label{sec:extended_features}
\begin{enumerate}
 \setcounter{enumi}{\theenumTemp}

\item \textbf{Father-Age-of-Death($v$)} - $v$'s father age of death. The formal Father-Age-of-Death($v$) definition is:
\[\mbox{Father-Age-of-Death}(v) := \mbox{Age-of-Death}(\mbox{Father}(v)), \]
where the function \textit{Father} returns the father vertex of $v$, if one exists. 
Namely, $Father(v) := u, \mbox{ where }  u \in V \land  \mbox{ gender}(u)=1  \land  \exists(v,u,Parent,t) \in E$.

\item \textbf{Mother-Age-of-Death($v$)} - $v$'s mother age of death. The formal Mother-Age-of-Death($v$) definition is:
\[\mbox{Mother-Age-of-Death}(v) := \mbox{Age-of-Death}(\mbox{Mother}(v)), \]
where the function \textit{Mother} returns the mother vertex of $v$, if one exists. 
Namely, $\mbox{Mother}(v) := u, \mbox{ where }  u \in V \land  \mbox{ gender}(u)=2  \land  \exists(v,u,Parent,t) \in E$.

\item \textbf{Paternal-Grandfather-Age-of-Death($v$)} - $v$'s paternal grandfather's age of death, if one exists. The formal Paternal-Grandfather-Age-of-Death($v$) definition is:
\[\mbox{Paternal-Grandfather-Age-of-Death}(v) := \mbox{Age-of-Death}(\mbox{Father}(\mbox{Father}(v))). \]

\item \textbf{Maternal-Grandfather-Age-of-Death($v$)} - $v$'s maternal grandfather's age of death, if one exists. The formal Maternal-Grandfather-Age-of-Death($v$) definition is:
\[\mbox{Maternal-Grandfather-Age-of-Death}(v) := \mbox{Age-of-Death}(\mbox{Father}(\mbox{Mother}(v))). \]

\item \textbf{Paternal-Grandmother-Age-of-Death($v$)} - $v$'s paternal grandmother's age of death, if one exists. The formal Paternal-Grandmother-Age-of-Death($v$) definition is:
\[\mbox{Paternal-Grandmother-Age-of-Death}(v) := \mbox{Age-of-Death}(\mbox{Mother}(\mbox{Father}(v))). \]

\item \textbf{Maternal-Grandmother-Age-of-Death($v$)} - $v$'s maternal grandmother's age of death, if one exists. The formal Maternal-Grandmother-Age-of-Death($v$) definition is:
\[\mbox{Maternal-Grandmother-Age-of-Death}(v) := \mbox{Age-of-Death}(\mbox{Mother}(\mbox{Mother}(v))). \]

\item \textbf{Sibling-Number($v$)} - the number of brothers and sisters $v$ had. The formal Sibling-Number($v$) definition is:
\[\mbox{Sibling-Number}(v) := |\{u \in V|u \in V \land  \exists(v,u,Sibling, t) \in E\}|.\]

\item \textbf{Max-Sibling-Age-of-Death($v$)} - the maximum age of death of $v$'s siblings. The formal Max-Sibling-Age-of-Death($v$) definition is:
\begin{multline*}
\mbox{Max-Sibling-Age-of-Death}(v) := \\ max(\{\mbox{Age-of-Death}(u) \neq \emptyset | u \in V \land  \exists(v,u,Sibling,t) \in E \}).
\end{multline*}

\item \textbf{Avg-Sibling-Age-of-Death($v$)} - the average age of death of $v$'s siblings. The formal Avg-Sibling-Age-of-Death($v$) definition is:
\begin{multline*}
\mbox{Avg-Sibling-Age-of-Death}(v) := \\ avg(\{\mbox{Age-of-Death}(u) \neq \emptyset | u \in V \land  \exists(v,u,Sibling,t) \in E\}).
\end{multline*}

\end{enumerate}

Using the features defined above, we specify the following feature sets, which will later be used to construct our multiple linear regression models and ML classifiers: (a) \textit{All-Numeric-Features} - a set which contains all the defined-above features that return numeric values, except the Death-Year feature;  (b) \textit{Heritage-Features} - a set which includes all the extended family  features, including the Birth-Year and Gender features; and (c) \textit{Nuclear-Family-Features} - a set which includes all the nuclear family features, including  Birth-Year and Gender. 

\subsection{Statistical and Predictive Analysis}
\label{sec:stat}
In this study, we used various algorithms and methods to calculate the variations in human lifespan over the past centuries, to identify which features are correlated with human lifespan and longevity, and to create predictive models which can assist in predicting human lifespan. 

In the remainder of this subsection, we describe in detail each one of our methods and algorithms.

\subsubsection{Lifespan Variations over Time}
\label{sec:variations}
After we had extracted the features for each vertex in the graph, we could utilize these features to calculate the variations in human lifespan over an extended period of time. To perform these calculations, we created two vertices datasets. The first dataset was the \textit{All-Dataset}, which included all the vertices with valid values of Age-of-Death, while the second dataset was the \textit{<Country>-Dataset} which included only vertices with valid values of Age-of-Death of people who were born in a specific country - in this study, we chose to take a closer look at people born in the United States. 

We utilized the \textit{All-Dataset} and the \textit{United-States-Dataset} to specifically  look at the lifespan of people who were born in each quarter of a century between 1650 and 1900. For each quarter of a century on each dataset, we calculated the Age-of-Death distribution of those people born in the chosen quarter. For example, in the second dataset, we had a group of \s{22,021} people who were born in the United States and lived between 1700 and 1724; we then calculated the percent of the population that died at each age between 0 and 122.\footnote{122 is the maximum confirmed human lifespan~\cite{maxlifespan}.} 

Additionally, for the \textit{All-Dataset-10} and for the \textit{United-States-Dataset-10}, and for each year from 1650 to 1900, we calculated both the average and median lifespans of the people who were born in each year and outlived the age of 10.  We also repeated these average and median calculations for each gender, using the \textit{Male-Dataset-10}, \textit{Female-Dataset-10}, 
\textit{Male-United-States-Dataset-10}, and \textit{Female-United-States-Dataset-10} datasets.

\subsubsection{Linear Regression}
\label{sec:linear}
One of the main goals of this study was to identify features which are correlated with lifespan and with longevity. To identify features correlated with an inheritance of human lifespan, we computed for each extended family  feature, which was defined in Section~\ref{sec:extended_features}, a simple linear regression $Y=\alpha +\beta X$, where $Y$ was set to be the Age-of-Death vector, and $X$ was set to be selected feature values.  For each feature we chose only  vertices from the \textit{<Feature>-Dataset-10}, in which both the Age-of-Death value was greater or equal ten\footnote{We chose to use  a minimum lifespan of 10 to avoid adding infant and child mortality, which might be misreported.} and the selected feature value existed.\footnote{For features such as Max-Sibling-Age-of-Death and Min-Spouse-Age-of-Death, which involved calculation of minimum, maximum or average, we ignored vertices with missing values, although by definition these features returned a valid  value of 0.} We then evaluated each simple linear regression  by computing the regression's \textit{P-value} and \textit{R-squared} values.

To identify if an individual's lifespan was correlated with the lifespan of his or her spouse(s), we repeated the same process of constructing a simple linear regression between the Age-of-Death feature and the Avg-Spouse-Age-of-Death, Max-Spouse-Age-of-Death, and Min-Spouse-Age-of-Death features. However, this time we used the \textit{Married-Dataset} to include only individuals who were married at least once.

To identify if longevity is correlated with reproductive success, we repeated the same process of constructing a simple linear regression between the Age-of-Death feature and the Children-Number feature. However,  with respect to Westendorp and Kirkwood's~\cite{westendorp1998human} results in mind, we used \textit{Children-Number-Dataset-50}, \textit{Male-Children-Number-Dataset-50}, and \textit{Female-Children-Number-Dataset-50} datasets, which only contained vertices with age of death of at least 50, namely after menopause.

\subsubsection{Multiple Linear Regression}
In this study, we used backward stepwise multiple linear regression to create models 
for predicting the Age-of-Death of individuals who had been born by 1900. We constructed these regression models by using the \textit{All-Numeric-Features}, the \textit{Heritage-Features}, and the \textit{Nuclear-Family-Features} sets, which were defined  at the end of Section~\ref{sec:features}. For constructing our first two models, we only used  vertices from the  \textit{No-Missing–Dataset-10} dataset with valid complete values, including defined gender values, for each selected features set of vertices who outlived the age of 10. Additionally, to prevent bias due to  the tendency of people to get married and have children in later stages of life, for the Nuclear-Family-Features set we only used vertices from the \textit{No-Missing–Dataset-50} dataset, i.e., those who outlived the age of fifty.

We evaluated these multiple linear regression models by calculating the \textit{P-value}, as well as the \textit{Multiple R-squared}, \textit{Adjusted R-squared}, and \textit{Residual Standard Error} (RSE) values.

\subsubsection{Machine Learning Algorithms}
\label{sec:ml}
One of the major drawbacks of using online genealogy datasets is the issue of missing values. In many genealogy datasets not all the profile data is complete;  many profiles contain missing values due to nonexistent data or privacy considerations~\cite{wikitree_privacy}. To overcome the issue of missing values and still gain predictive information from the profiles with nonexistent data, we chose to use Machine Learning algorithms, such as  decision trees and Naive-Bayes algorithms, which can deal with missing values. 

We evaluated various supervised learning algorithms in an attempt to predict which individuals who were born in the United States between 1650 and 1900, and outlived the age of fifty, will also outlive the age of 80.  We constructed our classifiers using Weka~\cite{Hall:2009:WDM:1656274.1656278}, a popular suite of ML,  and the features defined in the \textit{United-States–Dataset-50} dataset.  We used all numeric features in each dataset, except the Age-of-Death  and Death-Year features. Additionally, we also treated unknown gender values as missing values, instead of replacing them with 0 values.
Using these datasets as training sets, we used Weka’s OneR, C4.5 (J48) decision tree, K-Nearest-Neighbors (IBk;  with K=3,5), Naive-Bayes,  RandomForest, and Bagging implementations of the corresponding algorithms. 
For each of these algorithms, most of the configurable parameters were set to their default values except for the  J48 decision tree classifier, in which the pruning option was not enabled.
We evaluated each classifier using the 10-folds cross validation method and calculated the True-Positive, False-Positive, F-Measure, and the Area-Under-Curve (AUC) measure. The AUC is a standard way to compare classifier performances~\cite{bradley1997use}, in which 0.5 a value represents a random classifier.  

Additionally, to obtain an indication of the usefulness of the various features,  we analyzed their importance using Weka's information gain attribute selection algorithm.

\subsection{WikiTree Dataset}
\label{sec:wiki}
To test and evaluate our methods and algorithms, we chose to use information collected from the WikiTree  
website. This is a free and accessible collaborative family history website started by Chris Whitten~\cite{wikitree_about}, and it  contains more than 5 million profiles~\cite{wikitree5m} of individuals who primarily lived in the past. WikiTree contains many profile pages of people who lived in the previous centuries, and many of the profiles contain the following details  about each individual: full name, gender, date of birth, date of death, location of birth, location of death, parents' profiles, children's profiles, spouses' profiles, and siblings' profiles. Often, in order to maintain the privacy of still-living people, the website limits access to their profile personal details~\cite{wikitree_privacy}. In order to maintain the integrity of WikiTree profile data, many profiles give reference to the source of the data presented in the profile, and most profiles have a profile manager who has primary responsibility for WikiTree profiles~\cite{wikitree_manager}. In addition, to prevent editing of profiles by untrusted users, each WikiTree profile has an independent ``Trusted List'' of people who can edit and view the profile~\cite{wikitree_trust}, making the data in many profiles only editable to a limited number of people. 

To collect profile information from WikiTree, we developed a web crawler which crawled and parsed only public  profiles from the website.  Using our crawler, we have downloaded and parsed \s{1,070,189} public profile pages. 
Using these profiles, we were able to construct a directed multigraph with \s{9,192,212} links and \s{1,382,752} vertices, \s{out of which 118,590 vertices represented at least distinct 28,011 private profiles. 
Moreover, the constructed multigraph contained  at least 416,030 vertices represent individuals  born in the United States, according to their profile pages. These  vertices were connected by 5,168,275 links to other vertices in the multigraph (see Figure~\ref{fig:multi_graph}, and Tables~\ref{tab:wikitree_stat} and~\ref{tab:sample_size}).}


\begin{table}[htb]
\centering
\caption{WikiTree Dataset Statistics \label{tab:wikitree_stat}}{ 
\begin{center}
\includegraphics[width=\textwidth]{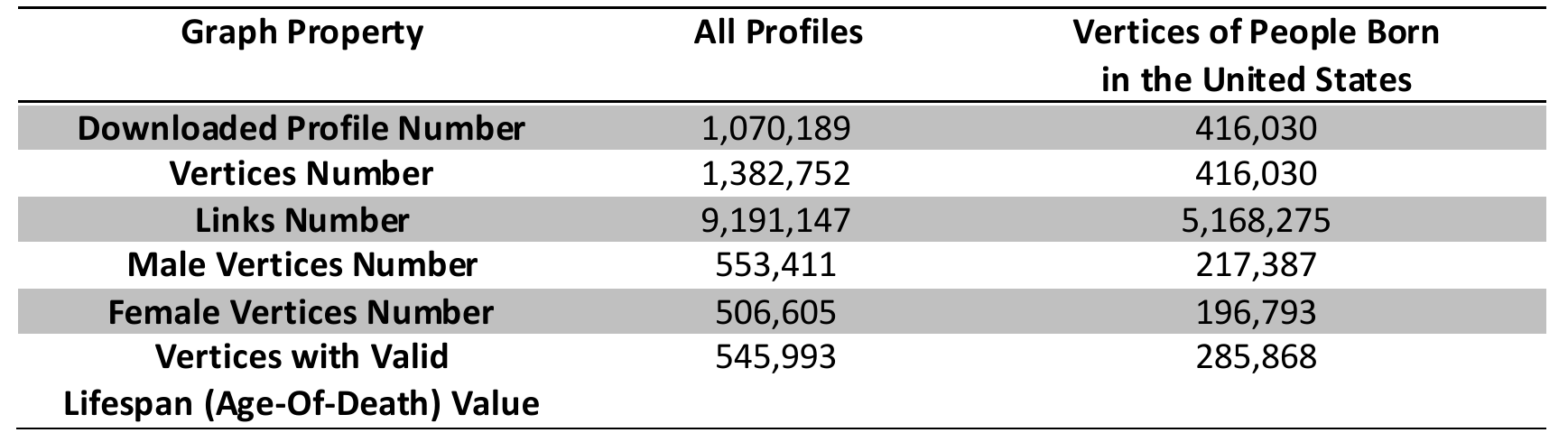}
\end{center}
}

\end{table}

\begin{table}[htb]
\centering
\caption{\textit{All-Dataset} Number of Profiles Born in Each Year  \label{tab:sample_size}}{ 
\begin{center}
\includegraphics[width=\textwidth]{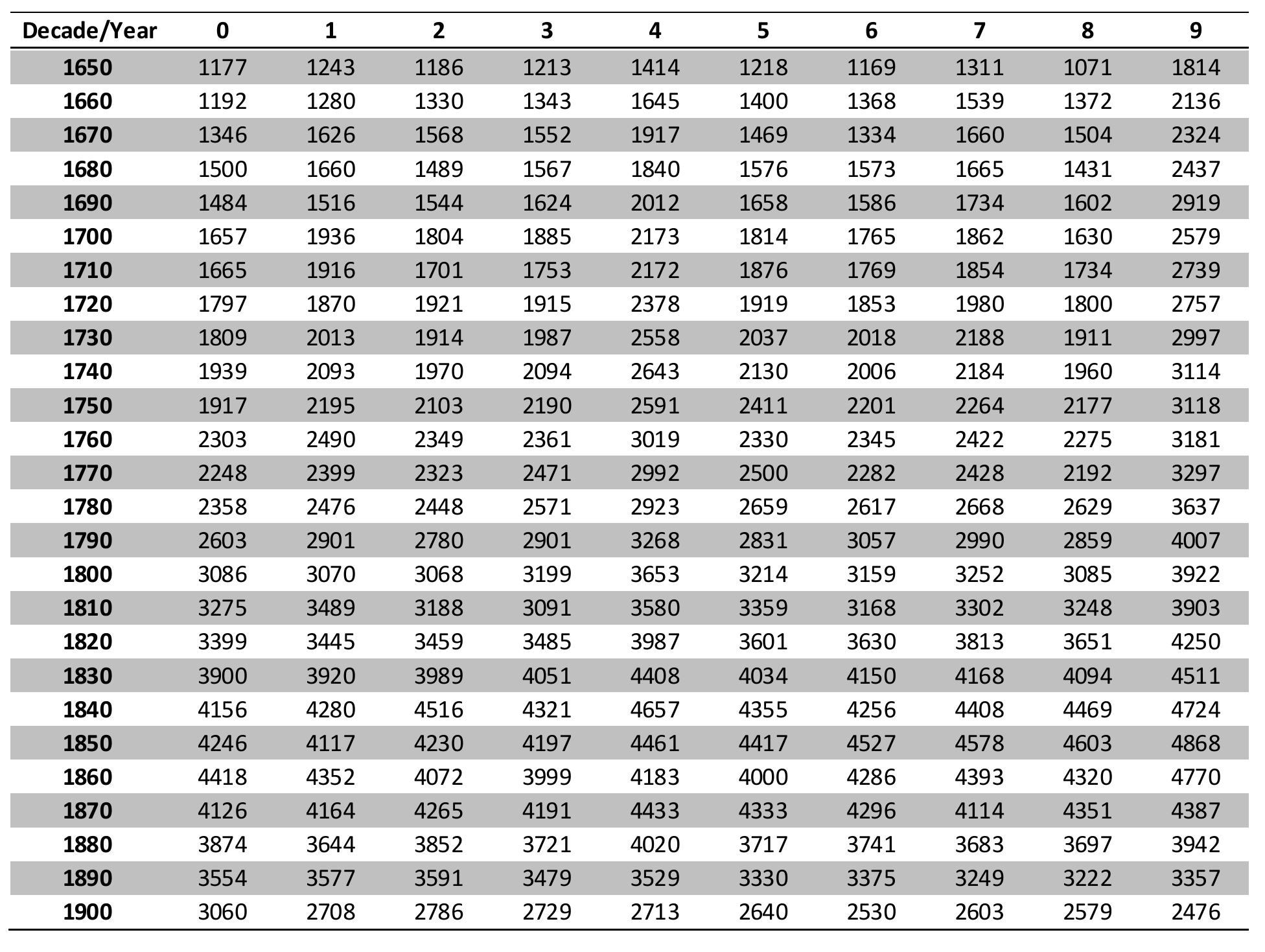}
\end{center}
}

\end{table}

\section{Results}
\label{sec:results}
In the following subsections, we present the results obtained using the algorithms and methods described in Section~\ref{sec:methods}. 
The results consist of three parts:  First, we present the results of calculating lifespan variations over time.  Second, we present the results of the simple linear regression and multi-linear regression analysis techniques  which were described in Section~\ref{sec:linear}. Finally, we present the results of the ML algorithms mentioned in Section~\ref{sec:ml}.

\subsection{Lifespan Variations over Time Results}
\label{sec:stat_results}
As described in Section~\ref{sec:variations}, we utilized the \textit{All-Dataset} and the \textit{United-States-Dataset} to compute the changes of lifespan over each quarter of a century between 1650 and 1900.
Then, we used these same datasets to take a closer look at the people who had been born during this 250-year span.  For each quarter of a century on each dataset, we calculated the Age-of-Death distribution of the people who were born in the chosen quarter. 
The results showing the lifespan variations over time for the \textit{All-Dataset} are presented in  Figure~\ref{fig:all_dataset_lifespan_var}, and for the \textit{United-States-Dataset}  in Figure~\ref{fig:us_dataset_lifespan_var}.

We also used the \textit{All-Dataset-10} and the \textit{United-States-All-Dataset-10} to calculate the average and the median lifespans for each gender, and for both genders, in each year between 1650 and 1900. The results of these calculations are presented in Figures~\ref{fig:lifespan_avg} and~\ref{fig:lifespan_med}.

\begin{landscape}
\begin{figure}
\centering
\begin{subfigure}{.8\textwidth}
  \centering
  \includegraphics[width=0.9\linewidth]{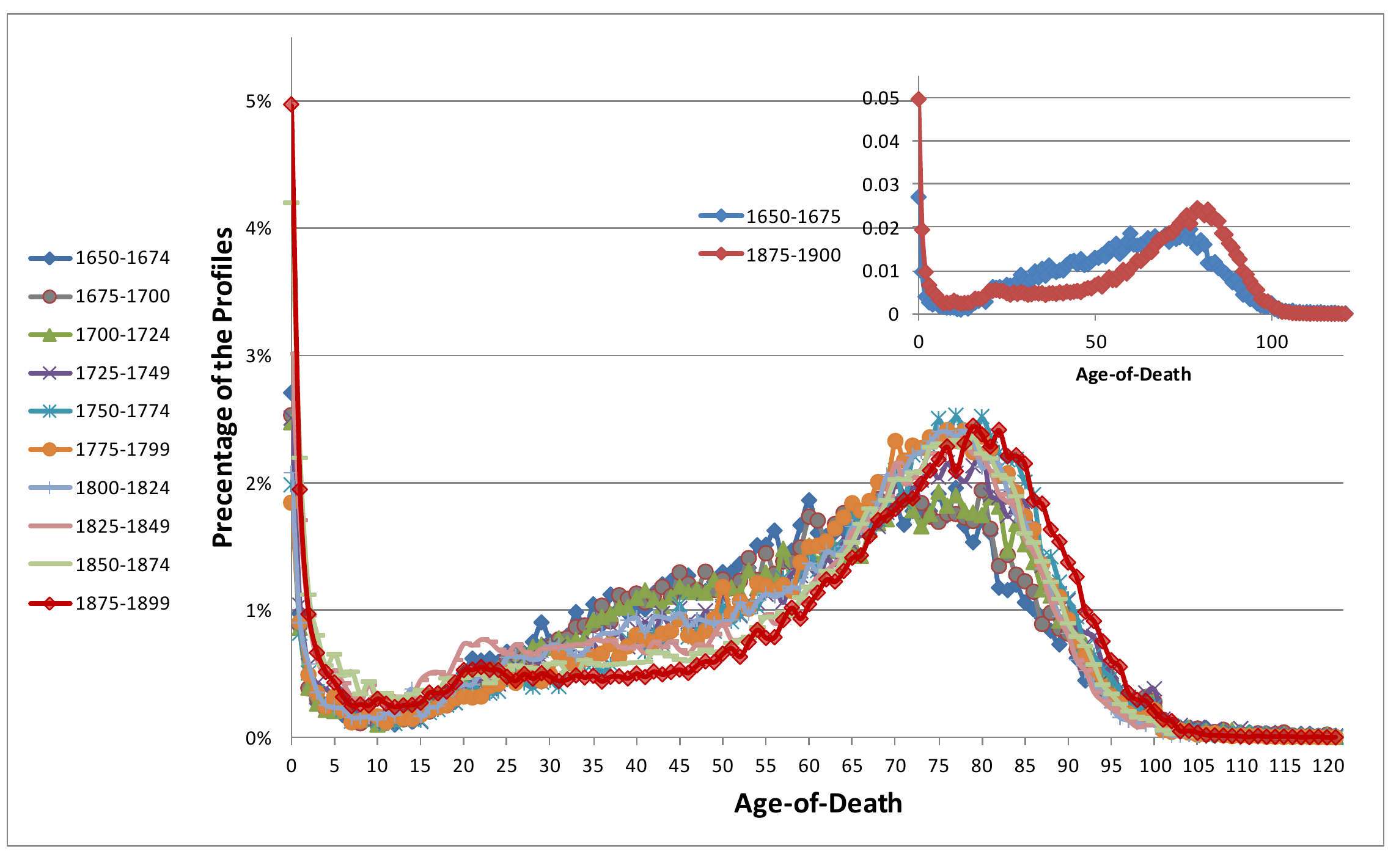}
  \caption{\textit{All-Dataset} Lifespan Variations.}
\label{fig:all_dataset_lifespan_var}
\end{subfigure}%
\begin{subfigure}{.8\textwidth}
  \centering
  \includegraphics[width=0.9\linewidth]{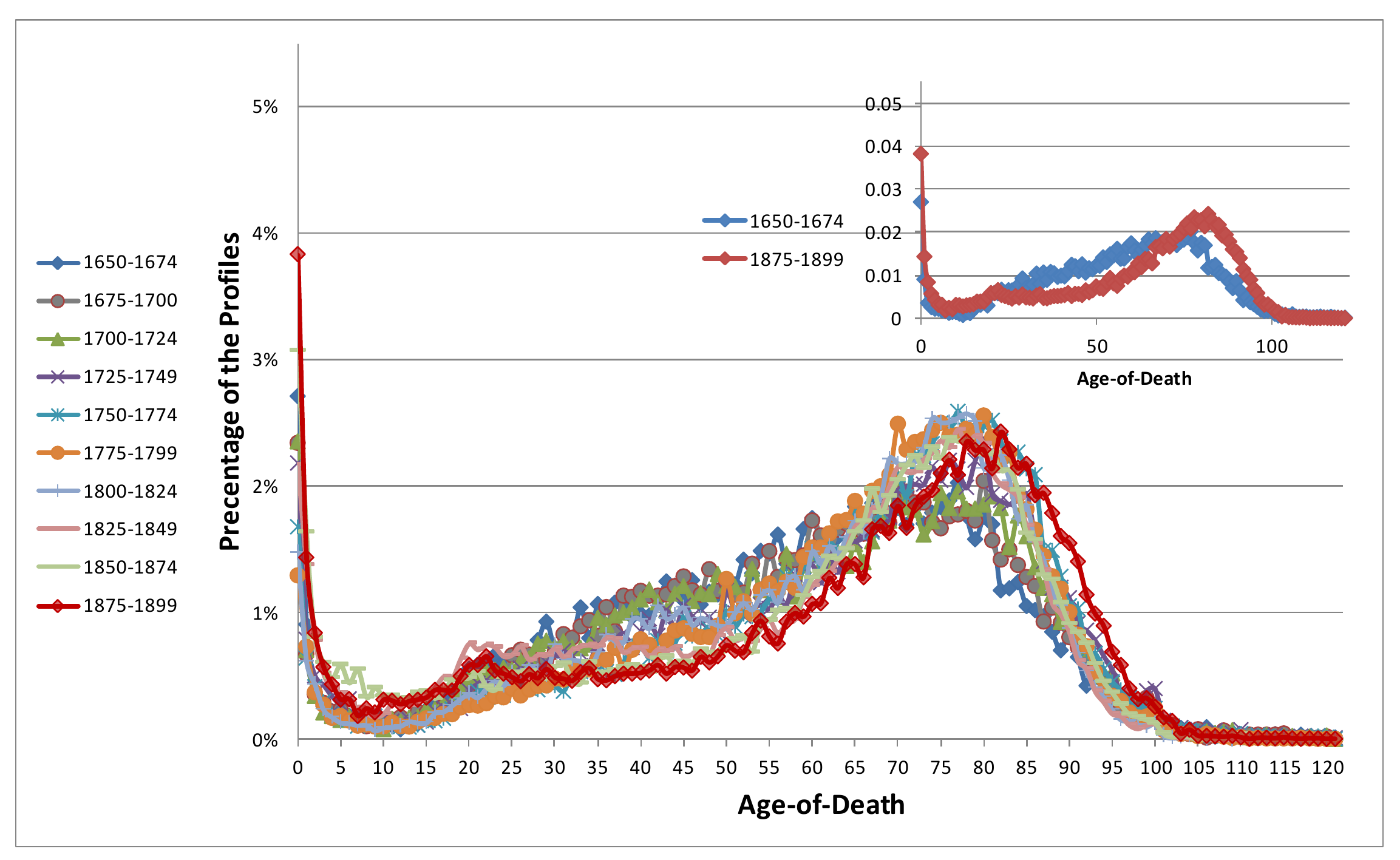}
  \caption{\textit{United-States-Dataset} Lifespan Variations. }
\label{fig:us_dataset_lifespan_var}
\end{subfigure}
\caption{Vertex Lifespan Variations, 1650-1900.}
\label{fig:lifespan_var}
\end{figure}

\begin{figure}
\centering
\begin{subfigure}{.8\textwidth}
  \centering
  \includegraphics[width=0.9\linewidth]{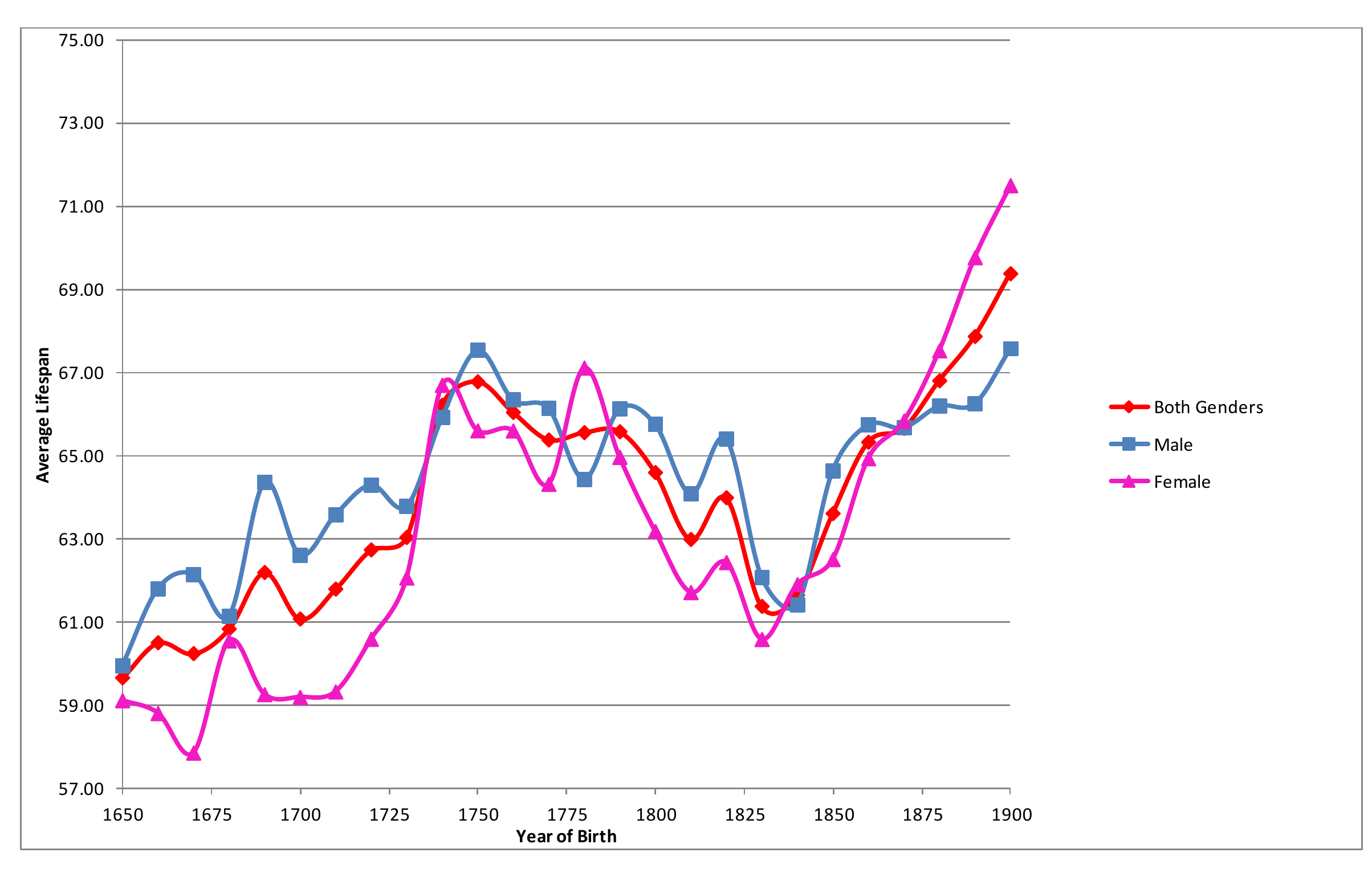}
  \caption{\textit{All-Dataset-10} - Average Lifespan. }
  \label{fig:all_avg}
\end{subfigure}%
\begin{subfigure}{.8\textwidth}
  \centering
  \includegraphics[width=0.9\linewidth]{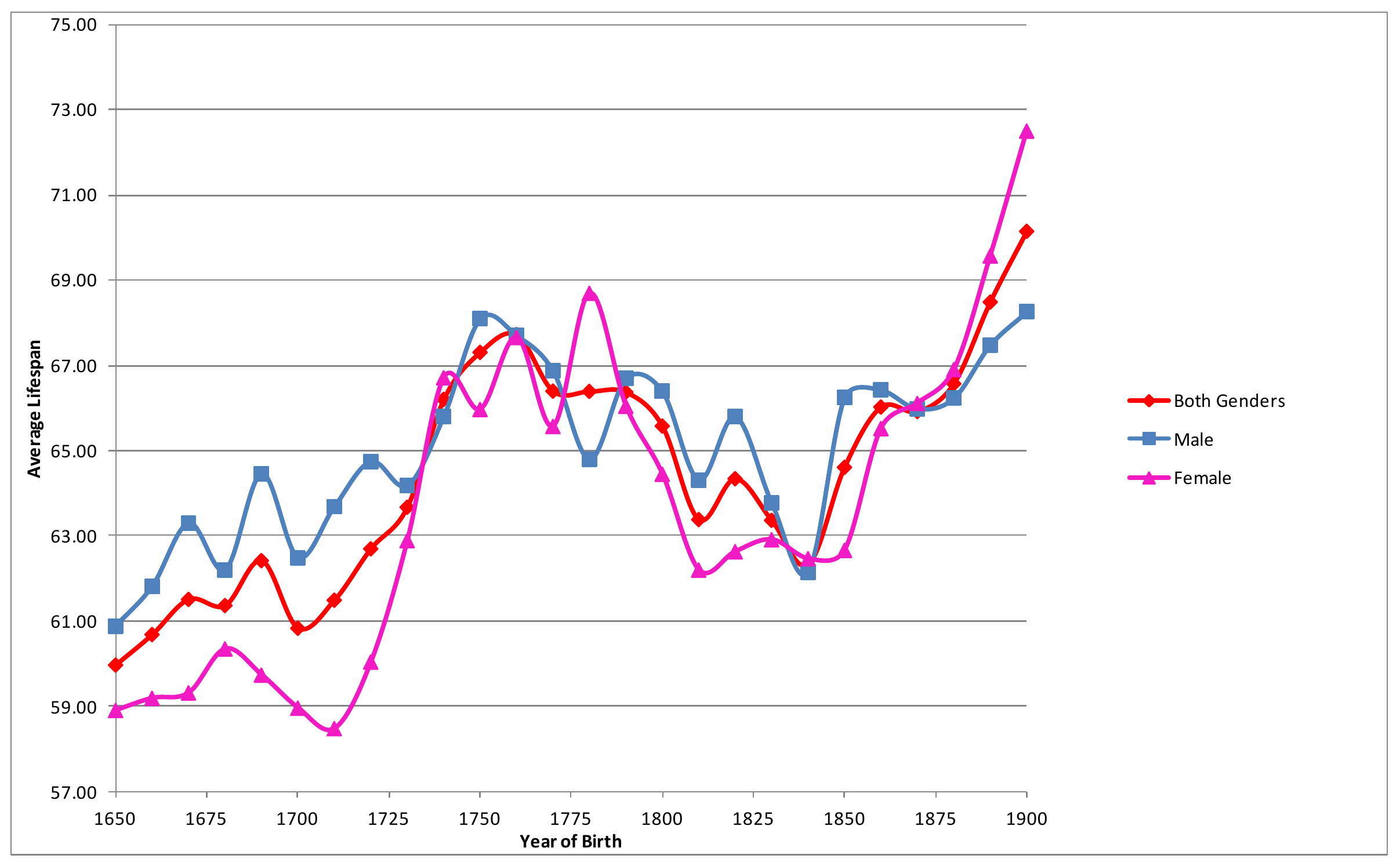}
\caption{\textit{United-States-Dataset-10} - Average Lifespan. }
  \label{fig:us_avg}
\end{subfigure}
\caption{Average Vertex Lifespans, 1650-1900. }
\label{fig:lifespan_avg}

\begin{subfigure}{.8\textwidth}
  \centering
  \includegraphics[width=0.9\linewidth]{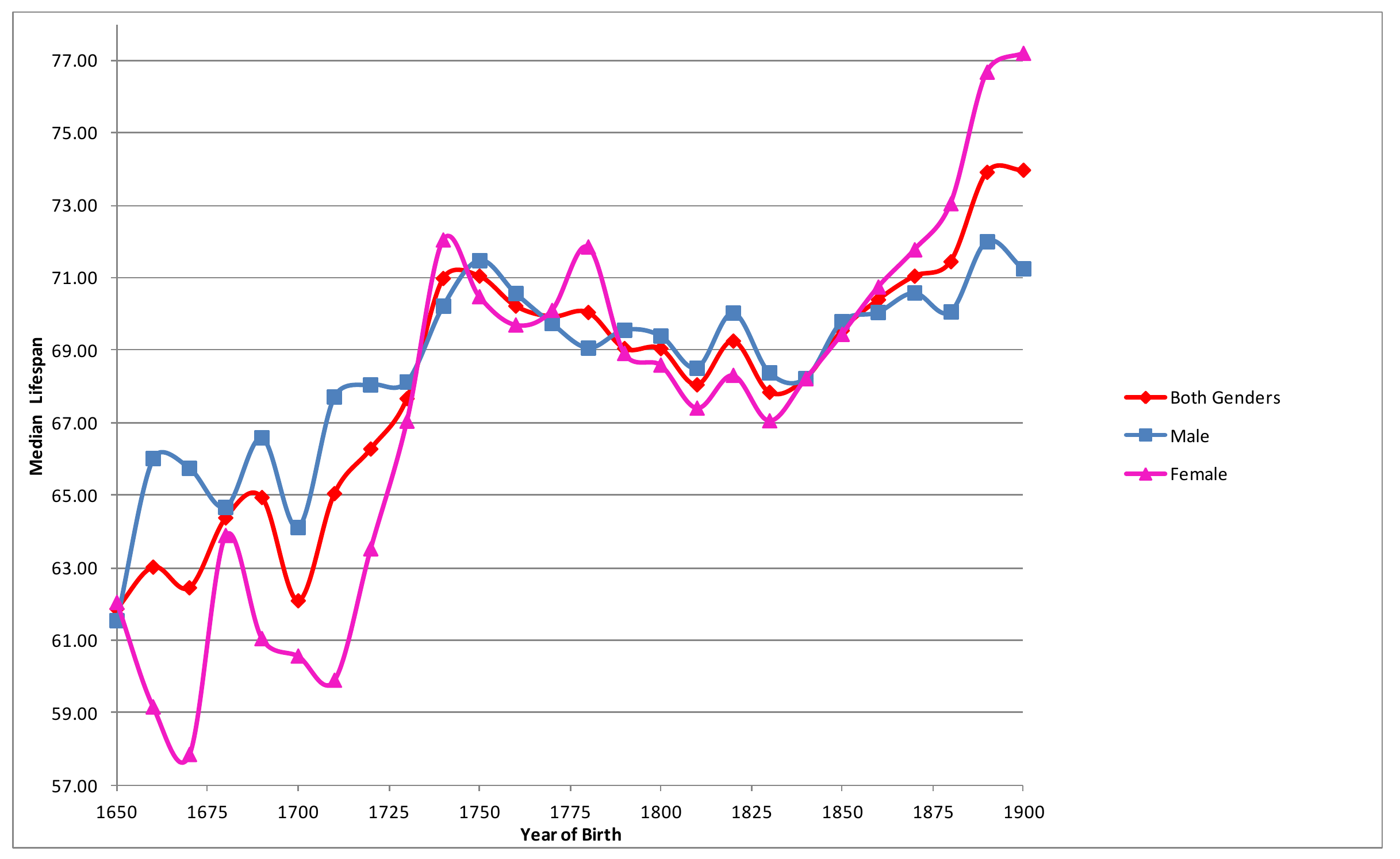}
  \caption{\textit{All-Dataset-10} - Median Lifespan.}
  \label{fig:all_median}
\end{subfigure}%
\begin{subfigure}{.8\textwidth}
  \centering
  \includegraphics[width=0.9\linewidth]{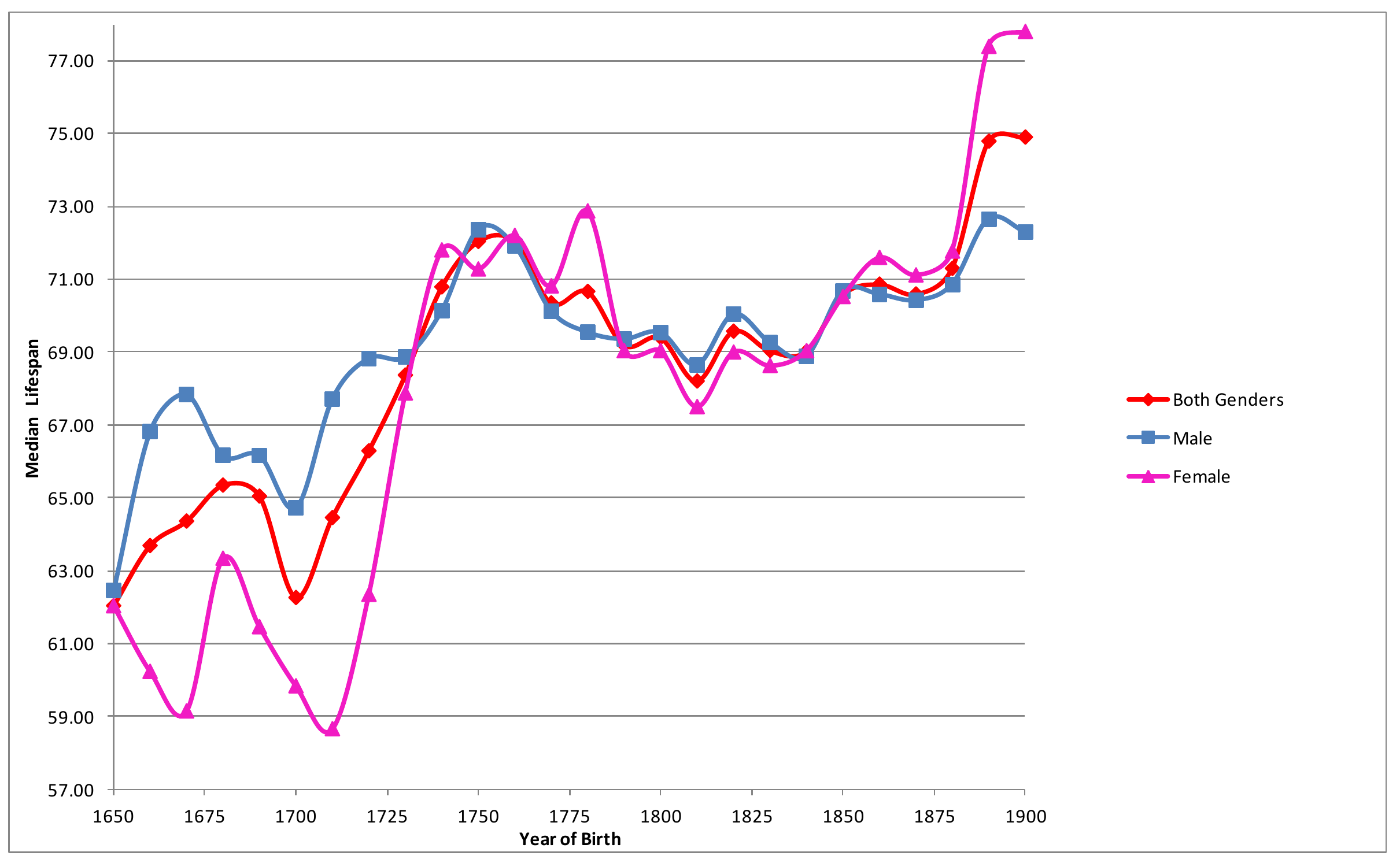}
\caption{\textit{United-States-Dataset-10} - Median Lifespan. }
  \label{fig:us_median}
\end{subfigure}
\caption{Median Vertex Lifespans, 1650-1900. }
\label{fig:lifespan_med}
\end{figure}
\end{landscape}
\subsection{Regression Analysis Results}
\label{sec:linear_results}
Using R-project software~\cite{team2005r}, we ran several simple linear and multi-linear regression algorithms based on the features we defined in Section~\ref{sec:features}. From the regression algorithms, we generated and evaluated several prediction models in order to determine the vertices' Age-of-Death. 

\subsubsection{Simple Linear Regression Analysis Results}
As described in Section~\ref{sec:linear}, we computed several simple linear regression models in order to predict linear correlations between the Age-of-Death and other features.
We first identified features correlated with the inheritance of human lifespan by computing  a simple regression model for each feature in the \textit{Extended Family Features} set, in order to predict the Age-of-Death feature.  In these calculations, we used  only the vertices who outlived the age of 10 and who also had valid existing information for each vertex's selected feature; i.e., we only used vertices which exist in the \textit{<Feature>-Dataset-10} for each selected feature. 

The simple regression results revealed that positive small but significant correlations  exist between most of the \textit{Extended Family Features} and the vertices' lifespans (see Table~\ref{tab:linear_results}). These correlations have R-squared values ranging from 0.0015 to 0.05, with a very low P-value of $2 \cdot 10^{-16}$ indicating that the correlation is highly significant, where the highest R-squared values were obtained for the Avg-Sibling-Age-of-Death (R-squared=0.05) and Max-Sibling-Age-of-Death (R-squared=\s{0.0272}) features, and the lowest R-squared values were obtained for the grandparents' lifespan features (R-squared ranging from 0.0015 to \s{0.0028}). Additionally, we also discovered a small negative correlation between the Sibling-Number feature and the vertices' lifespan, with a slope of \s{-0.155}, R-squared of \s{0.0021}, and P-value of 
 $2 \cdot 10^{-16}$.

We then repeated the simple linear regression calculation to identify correlations between the vertices' lifespans and their spouses' lifespans by using the Avg-Spouse-Age-of-Death, Max-Spouse-Age-of-Death, and Min-Spouse-Age-of-Death features with the \textit{Married-Dataset}. We discovered that each one of these features demonstrated a significant correlation, with a low P-value of $2 \cdot 10^{-16}$ and a maximum R-squared value of \s{0.0564} (see Table~\ref{tab:linear_results}).

\begin{table}[htb]
\centering
\caption{Simple Linear Regression Results \label{tab:linear_results}}{ 
\begin{center}
\includegraphics[width=\textwidth]{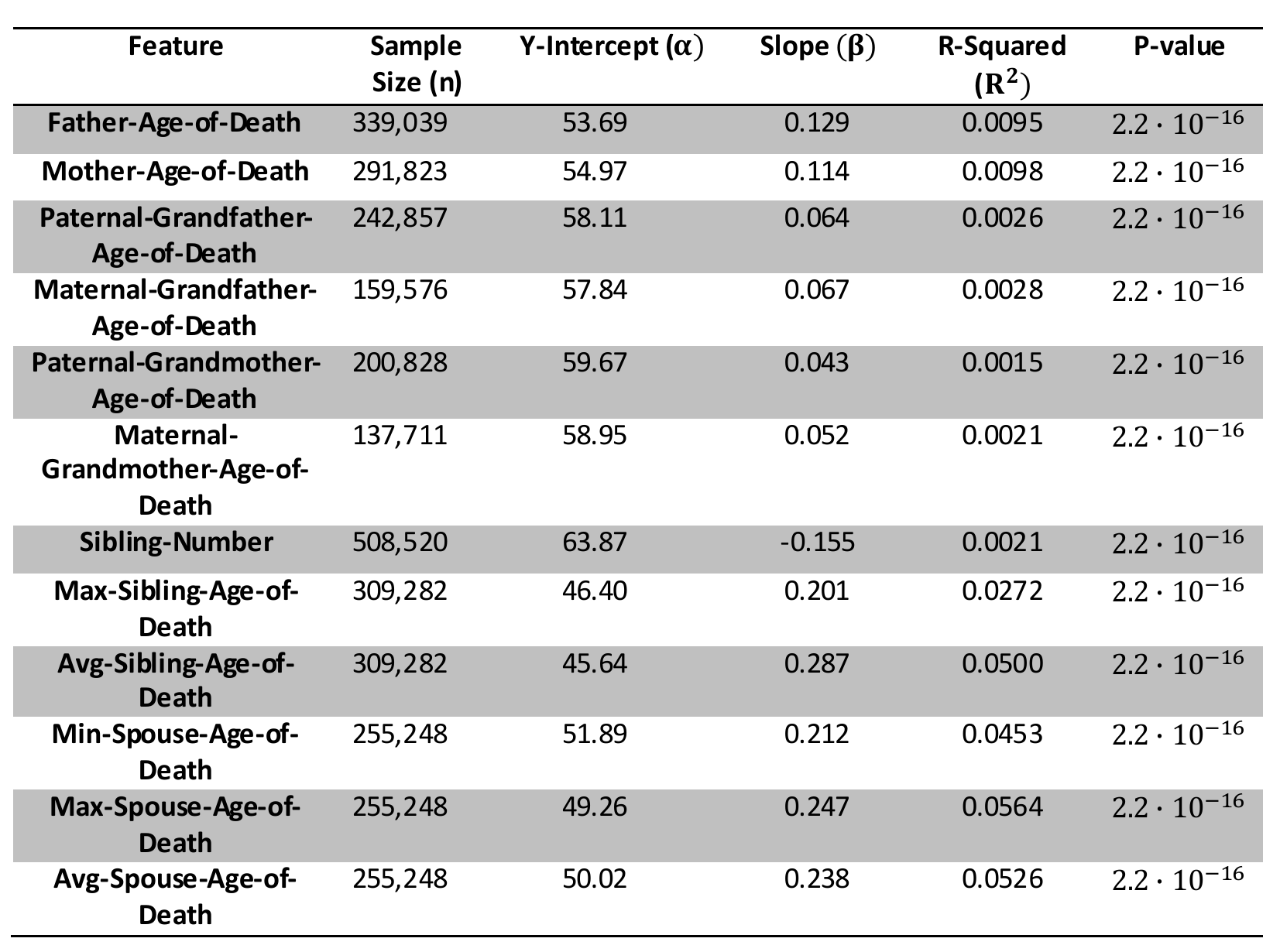}
\end{center}
}

\end{table}

Lastly, to identify if longevity is correlated with reproductive success, we computed simple linear regressions between the  Age-of-Death feature and the Children-Number feature on the following datasets: \textit{Children-Number-Dataset-50}, \textit{Male-Children-Number-Dataset-50}, and \textit{Female-Children-Number-Dataset-50}.
Using the simple linear regression, we obtained the following correlation results: (a) on the \textit{Children-Number-Dataset-50}  dataset ($n=375,938$) the regression returned a negative slope of -0.006, with a R-squared of $4.1\cdot 10^{-6}$ and a  \s{P-value of $0.2137$}; (b) on the \textit{Male-Children-Number-Dataset-50} dataset \s{($n=214,864$)} the regression returned a positive slope of \s{0.044}, with  a R-squared of \s{0.0002} and a \s{P-value of $1.64 \cdot 10^{-11}$}; and (c) on the \textit{Female-Children-Number-Dataset-50} dataset \s{($n=160,149$)} the regression returned a negative slope of \s{-0.079}, with a R-squared of \s{0.0006} and a P-value of  \s{$2.2 \cdot 10^{-16}$}.

\subsubsection{Multi-Linear Regression Analysis Results}
To create models which can estimate a vertex age of death based on the vertex's features, we chose to use the backward stepwise multiple linear regression technique. By combining this technique with the various predefined features sets, we created three multiple regression models which presented Multiple R-squared values of 0.085, 0.042, and 0.025, for the \textit{All-Numeric-Features} set, \textit{Heritage-Features} set, and \textit{Nuclear-Family-Features} sets respectively (see Table~\ref{tab:multi_linear_results}). 
We also obtained the following multiple linear regression models for each features set: For the 
\textit{All-Numeric-Features} set, we computed the following model using data collected from \s{59,893} vertices who were born by 1900 and outlived the age of 10:
\begin{eqnarray*}
\mbox{Age-of-Death}(v)& = &  9.813-2.194\cdot \mbox{Gender}(v) + \\
&& 0.023\cdot \mbox{Birth-Year}(v) + 
 0.261 \cdot \mbox{Children-Number}(v) + \\
&& 1.706 \cdot \mbox{Spouse-Number}(v) +  \\
&& 0.084\cdot\mbox{Max-Spouse-Age-of-Death}(v) + \\
&& 0.053\cdot\mbox{Father-Age-of-Death} + \\
&&0.039\cdot\mbox{Mother-Age-of-Death}(v) + \\
&& 0.015\cdot\mbox{Paternal-Grandfather-Age-of-Death}(v)  + \\
&& 0.016\cdot\mbox{Maternal-Grandfather-Age-of-Death}(v) +\\
&& 0.011\cdot\mbox{Paternal-Grandmother-Age-of-Death}(v)  + \\
&& 0.062 \cdot \mbox{Sibling-Number}(v) + \\
&& -0.1 \cdot \mbox{Max-Sibling-Age-of-Death}(v) +\\
&& 0.193 \cdot \mbox{Avg-Sibling-Age-of-Death}(v)
\end{eqnarray*}

For the \textit{Heritage-Features} set, we computed the following model using data collected from \s{59,893} vertices who were born by 1900 and outlived the age of 10:
\begin{eqnarray*}
\mbox{Age-of-Death}(v)& =&  18.184-2.06\cdot \mbox{Gender}(v) + \\
&& 0.021\cdot \mbox{Birth-Year}(v) + \\
&& 0.057\cdot\mbox{Father-Age-of-Death} + \\
&&0.041\cdot\mbox{Mother-Age-of-Death}(v) + \\
&& 0.013\cdot\mbox{Paternal-Grandfather-Age-of-Death}(v)  + \\
&& 0.016\cdot\mbox{Maternal-Grandfather-Age-of-Death}(v) +\\
&& 0.016\cdot\mbox{Paternal-Grandmother-Age-of-Death}(v)  + \\
&& -0.136 \cdot \mbox{Max-Sibling-Age-of-Death}(v) +\\
&& 0.202 \cdot \mbox{Avg-Sibling-Age-of-Death}(v)
\end{eqnarray*}

For the \textit{Nuclear-Family-Features} set, we computed the following model using data collected from \s{349,118} vertices who were born by 1900 and outlived the age of 50:
\begin{eqnarray*}
\mbox{Age-of-Death}(v)& =&  47.73+1.063\cdot \mbox{Gender}(v) + \\
&& 0.013\cdot \mbox{Birth-Year}(v) +  \\
&& -0.034 \cdot \mbox{Children-Number}(v) + \\
&& -0.22 \cdot \mbox{Spouse-Number}(v) + \\
&& -0.08\cdot\mbox{Min-Spouse-Age-of-Death}(v) + \\
&&  0.098\cdot\mbox{Avg-Spouse-Age-of-Death}(v) 
\end{eqnarray*}

\begin{table}[htb]
\centering
\caption{Multiple Linear Regression Results}{ 
\begin{center}
\includegraphics[width=\textwidth]{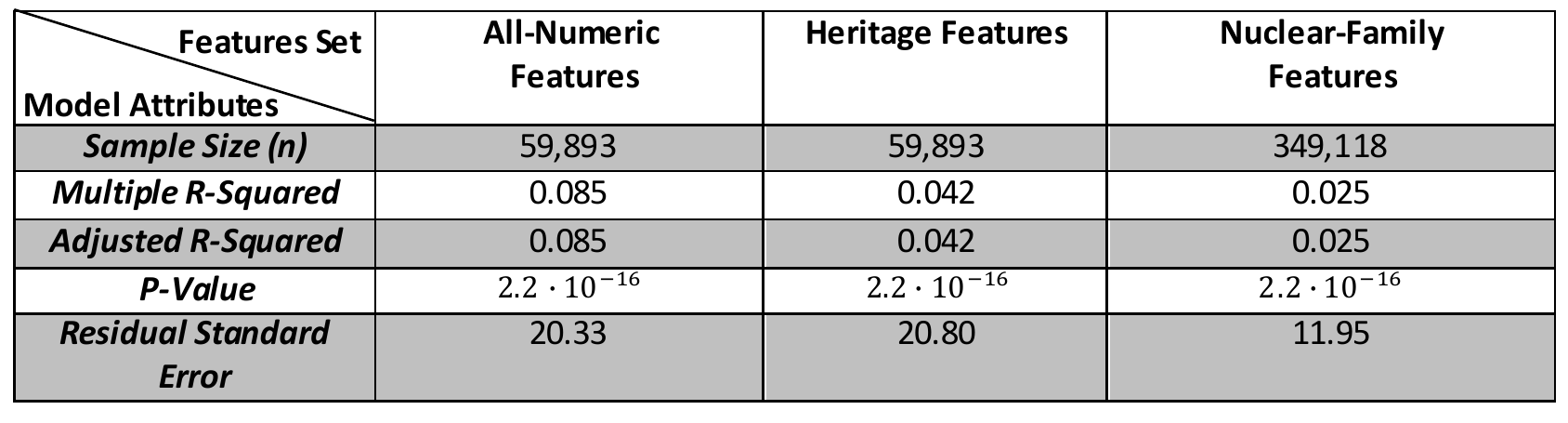}
\end{center}
}
\label{tab:multi_linear_results}
\end{table}

\subsection{Machine Learning Algorithms Results}
\label{sec:ml_results}
We evaluated various supervised learning algorithms in an attempt to predict which individuals who were born in the United States between 1650 and 1900 and outlived the age of fifty, will also outlive the age of eighty. 
We constructed our classifiers using the \textit{United-States–Dataset-50} dataset, which contained features of \s{183,494} vertices who outlived the age of fifty, out of which \s{58,975} vertices outlived the age of 80. 
To better understand which features were most useful to our classification algorithms, we analyzed the various features' importance using Weka's information gain features selection algorithm.
For the \textit{United-States–Dataset-50} dataset, 
the top eight features with the highest rank retrieved from Weka's information gain features selection algorithm were: \s{(a) Birth-Year (0.0058), (b) Max-Sibling-Age-of-Death (0.0047), (c) Avg-Sibling-Age-of-Death (0.0023), (d) Max-Spouse-Age-of-Death (0.0019), (e) Gender (0.0018), (f) Avg-Spouse-Age-of-Death (0.0016), (g) Min-Spouse-Age-of-Death (0.0016),
(h) Father-Age-of-Death (0.0008), (i) Mother-Age-of-Death (0.0006), and (j) Parental-Grandfather-Age-of-Death (0.0002). 
At the end of the list, the Maternal-Grandfather-Age-of-Death, the Children-Number, and the Sibling-Number features received an information gain score of 0.}


On this dataset, the RandomForest classifier received the maximum AUC of \s{0.632}, better than a random classifier with AUC of 0.5, while the maximum True-Positive value of \s{0.976} was obtained by the Decision-Tree classifier, and the minimum False-Positive of \s{0.774} was obtained by K-Nearest-Neighbors (K=3) classifier  (see Table~\ref{tab:ml_results}). 
We used T-tests with a significance of 0.05 to compare the AUC results of the RandomForest and the naive OneR classifiers.  According to the T-test result, RandomForest classifier performed better in terms of AUC, than the naive OneR classifier.

\begin{table}[htb]
\centering
\caption{Machine Learning Classifiers Results}{ 
\begin{center}
\includegraphics[width=\textwidth]{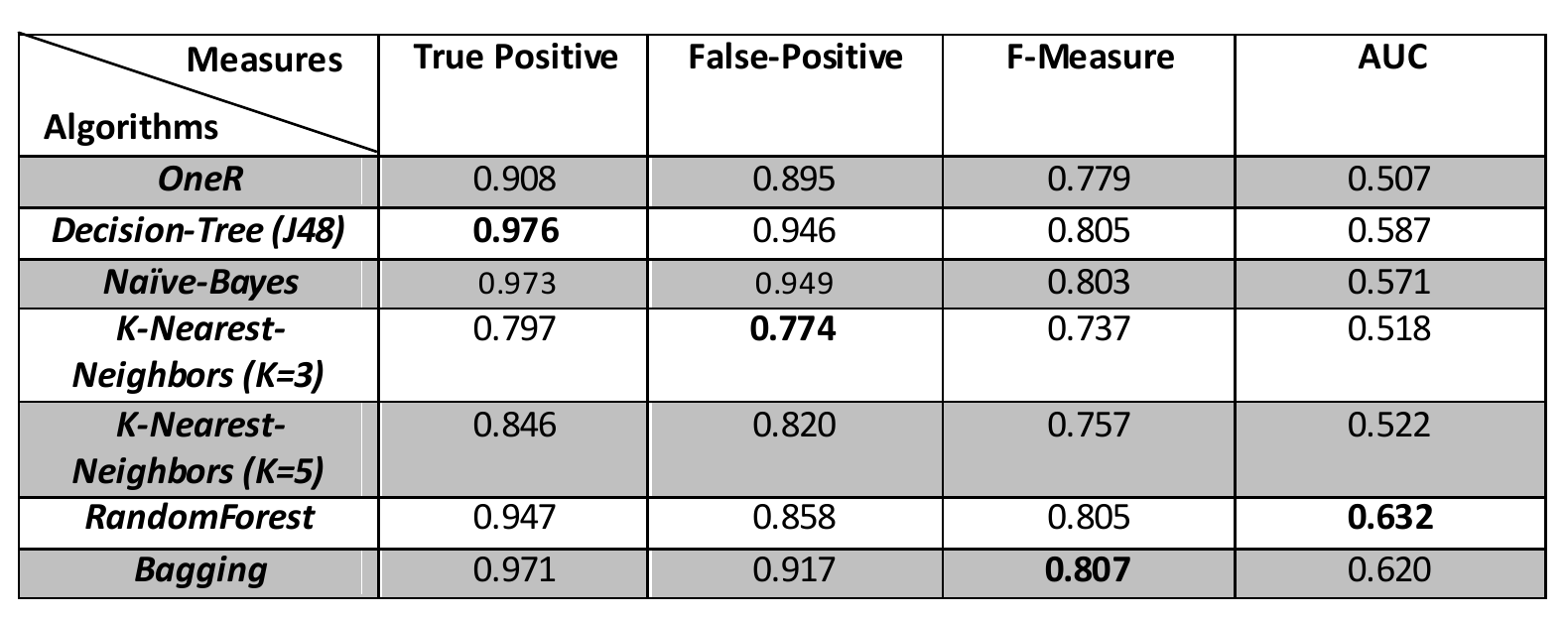}
\end{center}
}
\label{tab:ml_results}
\end{table}

\section{Discussion}
\label{sec:discussion}
To our knowledge, this study is the largest study to date  which utilizes genealogical datasets  to better understand factors that correlate with human lifespan. The algorithms and methods presented throughout this study, which were evaluated on the WikiTree dataset, reveal several interesting patterns and correlations.

Firstly, our results of lifespan variations over time, presented in Section~\ref{sec:stat_results} and in Figure~\ref{fig:lifespan_var}, demonstrate how the lifespans of human population changed over the previous centuries.
The lifespan graphs presented in Figure~\ref{fig:lifespan_var} show high infant and children death rates as well as local maximum values between the ages of 70 and 80; these results resemble the lifespan graphs  presented in Mitchell et al.~\cite{mitchell2001heritability} and by the UK Office of National Statistics~\cite{lifespan_graphs}. This resemblance supports our assumptions regarding the integrity of the WikiTree dataset, which indeed contains data on human population with largely accurate birth and death dates. However, the infant death rates presented in these graphs are not entirely accurate; according to Wegman~\cite{wegman2001infant}, in 1900 the infant mortality rates in the United States were about 15\%,  which is higher  than the values presented in our results. We assume that the main reason for this discrepancy was the lack of a uniform, formal definition of ``live births,'' which was not standardized until 1951~\cite{wegman2001infant}. Therefore, in most of this study we used as a sample set only people who outlived the age of ten.
Nevertheless, by analyzing these graphs, we can observe that over time, lifespans increased and fewer people passed away at young ages. 
Another observation that can be concluded from these graphs is that even in the second half of the seventeenth century, people who outlived the age of ten would likely outlive the age of sixty.  Indeed, according to our median lifespan analysis, presented in Figure~\ref{fig:us_median}, the median age in 1650  for people who were born in the United States and outlived the age of ten was \s{62.46} for males and \s{62.04} for females.

Secondly, our median and average population lifespan calculations, presented in Figures~\ref{fig:lifespan_avg} and~\ref{fig:lifespan_med}, reveal some interesting patterns. By analyzing the graphs, we can locate several years in which the average lifespans sharply decreased for  both males and females. For example, for people who were born in the United States in 1800 and outlived the age of 10, the average lifespans for males and females were \s{66.39} and \s{64.45}, respectively. However, for people born in the United States ten years later, in 1810, the average lifespan was reduced by around 2 years: males' lifespans decreased to \s{64.31}, and females' lifespans decreased to \s{62.20}. An additional and even more interesting reoccurring pattern can be identified in Figure~\ref{fig:us_median} in which, for a specific time period, the median lifespan for males increased while the median life span for females suddenly decreased, or vice versa. For example, from 1650 to 1660 the male median lifespan increased from \s{62.46}, to \s{66.82} while in the same period of time the female median lifespan decreased from \s{62.04} to \s{60.24}. Similar patterns  reoccur between 1770 and 1780, only this time the female average lifespan increased from \s{65.57} to \s{68.69}, while the male average lifespan decreased from \s{66.87} to \s{64.79} (see Figure~\ref{fig:us_avg}).
Another interesting pattern can be found between 1850 and 1900 where in just a half a century the female average lifespan sharply increased from \s{62.66} to \s{72.5}. We hope to discover  underlying reasons for these patterns in our future research.

Thirdly, using simple linear regression algorithms, we uncovered small but significant correlations between various features and the Age-of-Death feature which are presented in Table~\ref{tab:linear_results}. We found small positive significant correlations between the \textit{Extended Family Features} and the Age-of-Death feature. For all these correlations, R-squared values were small and ranged from  0.0015 to 0.05 with a P-value of $2.2 \cdot 10^{-16}$, and these may indicate that lifespan can ``run in the family.'' However, due to the small R-squared values in our results, we can conclude that the influence of inherited lifespan is limited and, in fact, negligible after more than one generation. Alternately, the observed correlation could be explained due to socioeconomic  reasons: ancestors with long lifespan might also indicate a higher socioeconomic status, which can be passed on to their offspring. We also found significant correlations between the Avg-Spouse-Age-of-Death, Max-Spouse-Age-of-Death, and Min-Spouse-Age-of-Death features and the Age-of-Death feature, with a low P-value and a maximum R-squared value \s{0.0564} (see Table~\ref{tab:linear_results}). This indicates that correlations between the lifespans of spouses exist, supporting the claims for the existence of the ``widow effect.'' We hope to confirm this observation in a future study by taking a closer look at the time intervals between the deaths of married couples. 
Using simple linear regression models, we also identified small significant correlations between longevity and reproductive success. Namely, we discovered  negligible negative correlation between females and their number of children (R-squared = \s{0.0006}), and negligible positive  correlation between males and their number of children (R-squared = \s{0.0002}). 

Fourthly, using multiple linear regression models, we were able to construct models which can predict a person's age of death using various features that were extracted from the WikiTree social network directed multigraph. Our models presented a low P-value of $2.2\cdot 10^{-16}$ with Adjusted R-squared of up to 0.085 (see Table~\ref{tab:multi_linear_results}), indicating that the extracted features can indeed assist in predicting a person Age-of-Death based on data which was extracted from the WikiTree dataset. However, the relative low Adjusted R-squared values indicate that other external factors are also responsible  for influencing an  individual's lifespan. We hope to test these assumptions in future studies by merging WikiTree genealogy datasets with other datasets that contain additional information about individuals' habits and lifestyles.

Fifthly, our machine learning classifiers presented better than random performances, with AUCs up to \s{0.632} (see Table~\ref{tab:ml_results}), in identifying which people who were born in the United States and outlived the age of 50 would also outlive the age of 80. These results support our previous claims that the data collected from genealogy datasets can be utilized to predict a person's lifespan. 
Additionally, the information gain algorithm results revealed that  the Max-Sibling-Age-of-Death, Avg-Sibling-Age-of-Death, and Max-Spouse-Age-of-Death (see Section~\ref{sec:ml_results}) were among the most useful features. These results also indicate that a correlation exists both between spouses' lifespans and between siblings' lifespans. In our future studies, we hope to use similar techniques to predict other personal attributes  based on data collected from online genealogy datasets.

The study presented here is among the first of its kind and offers many future research directions to pursue. One possible research direction is to analyze not only the structured data which appear in the WikiTree profile pages, but also to use Natural Language Processing (NLP) algorithms to analyze content data which appear in these pages. Another possible research direction is to compare the results presented in this study from the WikiTree dataset to other online genealogy datasets, such as FamiLinx,\footnote{\url{http://erlichlab.wi.mit.edu/familinx/data.html}} which is publicly available and contains information from Geni.com,\footnote{\url{http://www.geni.com/}}  a genealogy-driven social network.
Additionally, in future work,  we plan to utilize the results obtained through this study on the population of United States and test them on various populations in other countries.

\section*{Acknowledgment}
The authors would like to thank Carol Teegarden for her editing expertise and helpful advice.

\bibliographystyle{abbrv}
\bibliography{wikitree_lifespan}

\end{document}